\documentclass[twocolumn,prl,showpacs,superscriptaddress]{revtex4}
\usepackage{amssymb}
\usepackage{amsmath}
\usepackage{graphicx}
\usepackage{subfigure}
\usepackage{natbib}
\usepackage{epsfig}
\usepackage{amsfonts}
\usepackage{mathrsfs}
\usepackage{ulem}
\usepackage{color}
\usepackage[toc,page,title,titletoc,header]{appendix}
\begin{document}

\title{Phase diagram and topological superfluid state of spin-orbit coupled Fermi gas with attractive interactions in a one-dimensional optical lattice}
\author{Haiping Hu}
\affiliation{Beijing National Laboratory for Condensed Matter
Physics, Institute of Physics, Chinese Academy of Sciences,
Beijing 100190, China}
\author{Chen Cheng}
\affiliation{Beijing Computational Science Research Center, Beijing 100084, China}
\author{Yucheng Wang}
\affiliation{Beijing National Laboratory for Condensed Matter
Physics, Institute of Physics, Chinese Academy of Sciences,
Beijing 100190, China}
\author{Hong-Gang Luo}
\affiliation{Beijing Computational Science Research Center, Beijing 100084, China}
\affiliation{Center for Interdisciplinary Studies $\&$ Key Laboratory for
Magnetism and Magnetic Materials of the MoE, Lanzhou University, Lanzhou 730000, China}
\author{Shu Chen}
\thanks{Corresponding author, schen@aphy.iphy.ac.cn}
\affiliation{Beijing National
Laboratory for Condensed Matter Physics, Institute of Physics,
Chinese Academy of Sciences, Beijing 100190, China}
\affiliation{Collaborative Innovation Center of Quantum Matter, Beijing, China}

\begin{abstract}
Based on density matrix renormalization group method, we investigate the spin-orbit coupled Fermi gas with attractive interactions in one-dimensional optical lattice and present a complete phase diagram for a quarter-filling system with intermediate strong interactions. We unveil the exotic pairings induced by spin-orbit couplings and the phase transitions between different pairing states, and further demonstrate the existence of the long-sought topological superfluid state at moderate magnetic field and spin-orbit coupling strength. For the particle conserved system, this topological superfluid state can be fixed by its gapless single particle excitation, Luttinger parameter and non-trivial boundary effect. Our study removes out the widespread doubt on the existence of topological superfluid in one dimension and paves the way for simulating topological superfluid states in cold atom settings.
\end{abstract}

\maketitle

{\it Introduction.-}
Systems of ultracold atoms have provided ideal platforms for simulating intriguing physics in condensed matter physics for their extremely purity and highly controllability. Among the most interesting topics, the study of topologically nontrivial states, including topological insulators and topological superconductors (superfluids) \cite{review}, has challenged our understanding of classification of states of matter.
As the spin-orbit coupling (SOC) plays a crucial role in realizing non-trivial topological states,
recent experimental progress on generating synthetic gauge fields and SOC \cite{agf} in cold atomic gases \cite{yjlinuniform,yjlinmag,yjlinele,bosonsoc2,fermionsoc1,fermionsoc2,zhangjing2d} has made it a promising goal to directly simulating topological phases in cold atom settings.

A hallmark of the one-dimensional (1D) topological superfluids (TSF) is the emergence of boundary Majorana zero modes \cite{majorana}, which is believed to be a key ingredient for topological quantum computation \cite{topoquantum} and attracted intensive recent studies \cite{altman,liangjunjun,huhp,quchunlei,Goth,CChen,HHu,keselman,topodege1,topodege2}.
The mechanism for realizing the TSF in spin-orbit coupled systems roots in the induced effective p-wave pairings \cite{kitaev}, accompanied by the Zeeman field and pairing correlation, which comes from artificial mean-field treatment or proximity effect to the usual s-wave superconductors.
For a quasi-1D quantum system with SOC, it inevitably suffers from strong quantum fluctuation, which prevents the formation of true long-range superfluid order and leaves mean-field treatment questionable.
In a pioneering work \cite{altman} by Ruhman et. al., based on an effective field theory analysis, they discussed the possibility of realizing TSF in 1D Rashba-type SOC systems with attractive interactions and concluded that the topological degeneracy only appears in a configuration including at least two topological regions, which is notably different from the previous mean field studies. The Majorana-like quasi-zero modes are associated with interfaces between distinct phases that may form in different regions of the harmonic trap due to the spatial variation of the chemical potential. Another work regarding to the pairings of the spin-orbit coupled Fermi gas on a 1D optical lattice \cite{liangjunjun} has demonstrated the existence of exotic pairing states, including BCS pairing state, Fulde-Ferrell-Larkin-Ovchinnikov (FFLO) state, and mixed pairing state where BCS and FFLO correlations coexist. However, some basic questions, i.e., whether an isolated 1D uniform quantum wire can support TSF states and how to characterize these states, are yet to be answered.

In this work, based on density matrix renormalization group (DMRG) algorithm \cite{dmrg1,dmrg2}, we study the exotic pairing states and phase diagram of the attractive Fermi gas with synthetic SOC in a 1D optical lattice. The interplay between SOC, Zeeman field and attractive interaction induces a series of phase transitions between different pairing states which are characterized by pairing correlations in momentum space. The TSF phase can exist at moderate magnetic field and SOC strength with filling factor deviating from the half-filling. The pairing of this topological phase in momentum space exhibits two sub-peaks and one main zero peak. For this non-trivial topological phase, the single-particle excitations are gapless. Under open boundary conditions (OBC), its transverse spin polarization, together with the single-particle excitation, mainly resides at two ends of the system. These results confirm the existence of the long-sought topological superfluid state.

{\it Model.-}We consider the Fermi gas with synthetic SOC trapped in a deep 1D optical lattice described by the following effective model
\begin{eqnarray}
H=H_{t}+H_{SOC}+H_{Z}+H_{U}
\end{eqnarray}
with
\begin{eqnarray}
H_{t}&=&-t\sum_{i,\sigma}(c_{i\sigma}^{\dag}c_{i+1\sigma}+h.c.),\\
H_{SOC}&=&\alpha\sum_i(c_{i\uparrow}^{\dag}c_{i+1\downarrow}-c_{i\downarrow}^{\dag}c_{i+1\uparrow}+h.c.), \\
H_{Z}+H_{U} &=& h\sum_i(n_{i\uparrow}-n_{i\downarrow})+U\sum_i n_{i\uparrow}n_{i\downarrow},
\end{eqnarray}
where $c_{i,\sigma}$ denotes the fermion annihilation operator with spin $\sigma$ at site $i$. The SOC term represents the hopping between nearest sites accompanied by spin flipping with strength $\alpha$. The strength of the on-site interaction $U$ can be fine tuned by Feshbach resonance or confinement induced resonance \cite{CIR}, $h$ is the strength of magnetic field, and the filling factor is defined as $\nu=N/L$. The above model has the particle-hole symmetry which is invariant under the transformation \cite{liangjunjun} $c_{i,\sigma}\rightarrow\sigma(-1)^{i}d_{i,-\sigma}^{\dag}$. Only filling $\nu\leq 1$ need to be considered. For convenience, we set $t=1$ as units of energy and take a fixed $U=-4$ in the subsequent calculations. Our DMRG calculations are performed under OBC with truncation error smaller than $10^{-5}$ for the ground state energy. 

To understand the possible exotic pairings, we briefly discuss the single particle spectrum $\epsilon(k)=-2t\cos k\pm\sqrt{4\alpha^2 (\sin k)^2+h^2}$, corresponding to the noninteracting Hamiltonian $H_0(k) = -2t \cos k I +  h \sigma_z - 2\alpha \sin k \sigma_y$ in the momentum space. In the absence of SOC, the magnetic field splits the spectrum into two bands with each band fully polarized along the opposite direction of $\sigma_z$. As shown in Fig.1(a), only interband finite-momentum pairings are possible if the attractive interaction is considered. On the other hand, if only SOC exists, there exists an effective momentum-dependent field along the $\sigma_y$ direction and the spectrum is also split into two bands as shown in Fig.1(b), with momentum-dependent polarization in each band.
For $k>0$, the polarization of the upper and lower band is $|\rightarrow \rangle$ and $|\leftarrow \rangle$, respectively, while for $k<0$, it is reversed. Once the attractive interaction is imposed, intra-band BCS-type pairings occur. Coexistence of the SOC and magnetic field opens a gap at $k=0$, and the phase transition between BCS and FFLO pairing states may be realized by tuning the SOC or magnetic field. As shown in Fig.1(c), when the filling $\nu=1$, both intra-band and inter-band pairings can be induced as the spin polarization contains both up and down components. This indicates the possibility of the coexistence of FFLO and BCS type pairings. For lower (or higher) fillings  as indicated by black solid lines, only intra-band pairings are possible. In the mean-field scheme, the system may enter a TSF phase with the existence of majorana zero modes at two ends of the chain.

\begin{figure}
\includegraphics[width=3.7in]{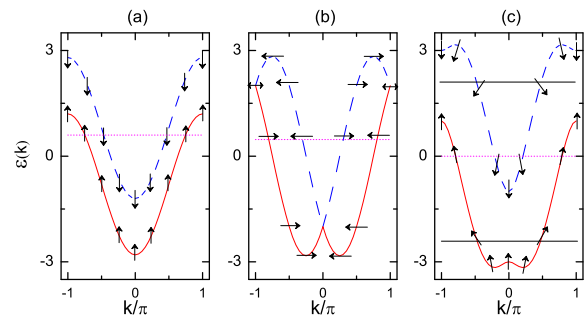}
\caption{(Color online) Dispersion relations of SOC Fermi gas with (a) only magnetic field; (b) only SOC; (c) both magnetic field and SOC. The red solid and blue dashed curves represent lower and upper bands, respectively. The arrows indicate the spin-polarizations in $\sigma_z$ and $\sigma_y$ directions. The horizontal dotted lines denote chemical potential. The black solid horizontal line corresponds to TSF permitted regions from mean-field results.}
\end{figure}
\begin{figure}
\includegraphics[width=3.7in]{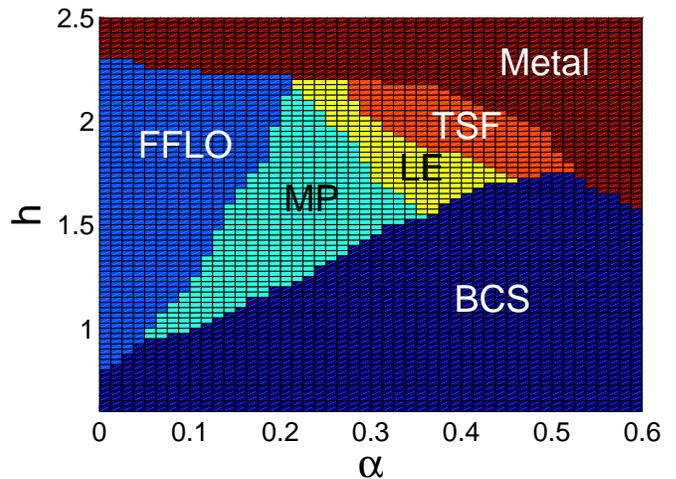}
\caption{(Color online) Phase diagram for the system with $N=24$ and $L=48$ determined by DMRG. The phase diagram contains six phases: BCS pairing phase, FFLO pairing state, mixed pairing state(MP), LE state, TSF state and metal state.}
\end{figure}

{\it Phase diagram and analysis of DMRG results.-} From the mean-field study \cite{huhp,quchunlei}, the topological phase should exist at filling deviating from $\nu=1$ and with a moderate magnetic field to insure the chemical potential lies in the Zeeman gap. As the system has the particle-hole symmetry, we here only consider the $\nu=1/2$ case. Before going into the detailed analysis of our DMRG results, we first summary the phase diagram in Fig.2 for a system with the lattice size $L=48$. There exist six phases in the $\alpha-h$ phase space, namely, BCS pairing phase, FFLO pairing state, mixed pairing state, Luther-Emery (LE) state, TSF state as well as metallic state. In order to identify different pairing states, we define the following two order parameters, i.e.,
the magnetization
\begin{equation}
\langle \sigma_z \rangle =\sum_i \langle n_{i\uparrow}-n_{i\downarrow} \rangle,
\end{equation}
and the singlet pairing
\begin{equation}
P_s^{i j}=\langle c_{i \downarrow}^{\dag} c_{i \uparrow}^{\dag} c_{j \uparrow} c_{j \downarrow} \rangle.
\end{equation}

The Fourier transformation of the real space pair correlation function $P_s(k) = \sum_{ij}P_s^{ij}e^{i k(i-j)}$ gives the pairing pattern in momentum space. For a small magnetic field, the system is always in the BCS pairing state. If the SOC strength is small, increasing magnetic field will induce a phase transition directly from the BCS state to the FFLO state \cite{FFLO1,FFLO2}, and then to the metal state. Things will largely change if the SOC strength is set at the moderate strength. The system undergoes first from the BCS state to the mixed pairing state \cite{liangjunjun} then to the FFLO state or the metal state. The BCS, FFLO and mixed pairing phase can be distinguished by the s-wave pairing correlations in momentum space. For the BCS phase, the pairing peak occurs at $k=0$ and the pairing correlation in real space is always positive. For the FFLO state, the pairing peak is at finite momentum $k=\pm Q$. The pairing correlation in real space oscillates around zero with period $1/Q$. For the mixed pairing state, the pairing peaks in momentum space appear at both $k=0$ and $k=\pm Q$. This novel three-peak structure comes from the interplay between SOC and Zeeman field. The most interesting thing in the phase diagram is the existence of a TSF phase if further increasing the SOC strength. In this case, sweeping magnetic field will induce a series of phase transitions: from the normal BCS state to the LE state \cite{LE,yhchan} where the charge excitation is gapless, then to the TSF state and metal state. In all the above phases, the single-particle excitations for LE, TSF and metal phases are gapless. The phase boundary between the metal phase and other phases can be determined by the Luttinger parameter $K_{\rho}$, which can be easily produced by calculating the density-density correlations in long-wave limit \cite{luttinger1,luttinger2,chengchen}, i.e.,
\begin{eqnarray}
K_{\rho}=\frac{N(k)}{k/\pi}, ~~~ k\rightarrow 0
\end{eqnarray}
where $N(k)=\sum_{j,l}e^{i k(j-l)}N_{jl}$, and $N_{jl}=\langle n_j n_l \rangle - \langle n_j\rangle \langle n_l \rangle$ is the density-density correlation function.

In Kitaev's original paper \cite{kitaev} for the topological p-wave superconducting chains, the $U(1)$ symmetry of particle number conservation is destroyed. The basic question is, in a charge conserved system, how can we determine whether a phase belongs to topological phase? The topological degeneracy \cite{topodege1,topodege2} which comes from different parity space will disappear as different parity states must have different particle numbers. What is worse, the low-energy excitation, i.e., the usual phonon excitation in one dimension from bosonization analysis, is proportional to $1/L$ which makes the assumed zero modes indistinguishable \cite{altman}. We define the so-called pair binding energy under OBC as follows:
\begin{eqnarray}
E_b=1/2[E_0(N-1)+E_0(N+1)]-E_0(N)
\end{eqnarray}
where $E_0(N)$ denotes the ground state energy with the total particle number $N$. When $N$ is an odd number, for the normal superfluid state, including FFLO, BCS, and mixed pairing phase, $E_b$ is finite but negative representing the particles are paired together. For an even $N$, it is positive and tends to a finite value in thermodynamical limit. For the metal phase, this quantity, which equals to the single-particle gap, is positive and tends to zero in thermodynamical limit. For the normal superfluid state, adding a particle into the system requires a finite amount of energy, whereas for the TSF state, the particle can be added at the end of the chain, which costs charging energy only \cite{altman,keselman} in a charge conservation system. In Fig.3, we do finite size analysis for the pair binding energy $E_b$ and make a linear fitting of the DMRG data. For the normal superfluid phase, $E_b$ is finite in the thermodynamical limit, while it tends to zero linearly for the metal phase. For both the LE and TSF phase, this quantity is close to zero in the thermodynamical limit.
\begin{figure}
\includegraphics[width=3.7in]{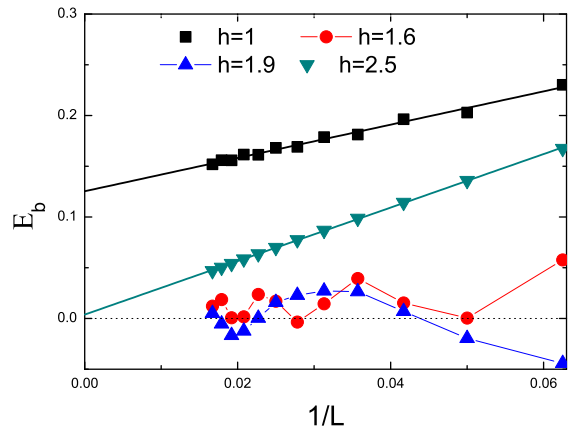}
\caption{(Color online) Finite size analysis of the pair binding energy $E_b$ for BCS state with $\alpha=0.35$, $h=1$; LE phase with $\alpha=0.35$, $h=1.6$; TSF phase with $\alpha=0.35$, $h=1.9$; metal phase with $\alpha=0.35$, $h=2.5$. The straight lines come from linear fitting. }
\end{figure}

Next we discuss the phase transitions between different phases in the phase diagram with the magnetic field fixed at $h=1.8$. By sweeping the SOC strength, the system undergoes five phases, i.e., the FFLO, mixed pairing phase, LE, TSF and metal phase by increasing SOC strength from $0$ to $0.6$. Fig.4 shows the main results. Fig.4(a) to (b) demonstrate the magnetization and pairing peak at $k=0$ and $k=k_F=\pi/2$ with respect to the SOC strength. We can clearly see four phase transitions: first from the FFLO state to the mixed pairing state at about $\alpha=0.18$, then to the LE state at about $\alpha=0.31$, to the TSF state at about $\alpha=0.42$, and metal state at about $\alpha=0.52$. The phase boundary can be determined from either the discontinuity of the slope of magnetization $\langle \sigma_z \rangle$ or pairing correlations in momentum space. In Fig.4(c), we show the Luttinger parameter $K_{\rho}$ with respect to the SOC strength. For the normal superfluid state, LE and TSF state, $K_{\rho}>1$, while for the metal state, $K_{\rho}<1$. The phase transition point determined via this way coincides with the result fixed by the order parameter before. In Fig.4(d), we give the pair correlations in momentum space. For $\alpha=0.1$, the system is in the FFLO phase with two pairing peaks located at some finite momentums. With the increase of the SOC strength, the $k=0$ peak emerges. For $\alpha=0.25$, the exotic three-peak structure appears \cite{liangjunjun}, corresponding to the so-called mixed pairing phase. Further increasing $\alpha$, two peaks at finite momentums are suppressed. At $\alpha=0.35$ and  $\alpha=0.46$, where the system is in LE and TSF phase respectively, the pair correlations in momentum space exhibit one main peak at $k=0$ and two sub-peaks at some finite momentums. Further increasing the SOC strength, the two sub-peaks disappear and the system enters into the metal phase with $K_{\rho}<1$.
\begin{figure}
\includegraphics[width=3.7in]{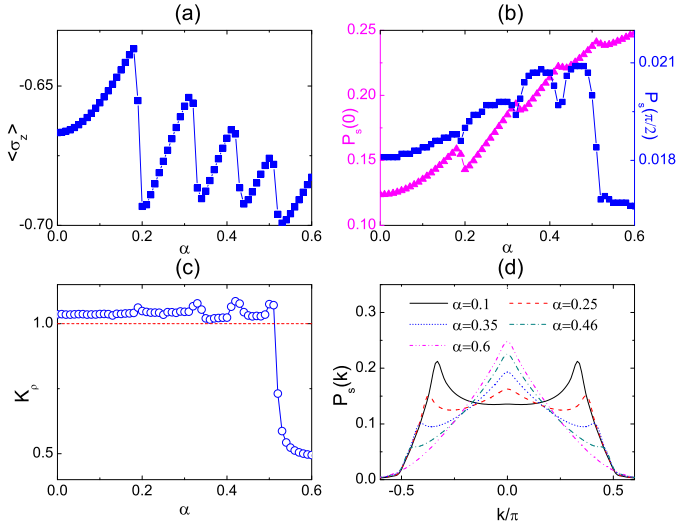}
\caption{(Color online) Phase transitions between different pairing states driven by the SOC for the system with $h=1.8$, $N=24$ and $L=48$. (a) The magnetization; (b) pairing correlations in momentum space: $P_s(0)$ and $P_s(k_F)$; (c) the Luttinger parameter $K_{\rho}$ versus $\alpha$. The red dotted line marks $K_{\rho}=1$ which serves as a criteria to distinguish the superfluid state and metal state. (d) Singlet pairings in momentum space for $\alpha=0.1$, $0.25$, $0.35$, $0.46$ and $0.6$, which correspond to the FFLO, mixed pairing state, LE, TSF, and metal phases, respectively.}
\end{figure}
\begin{figure}
\includegraphics[width=3.7in]{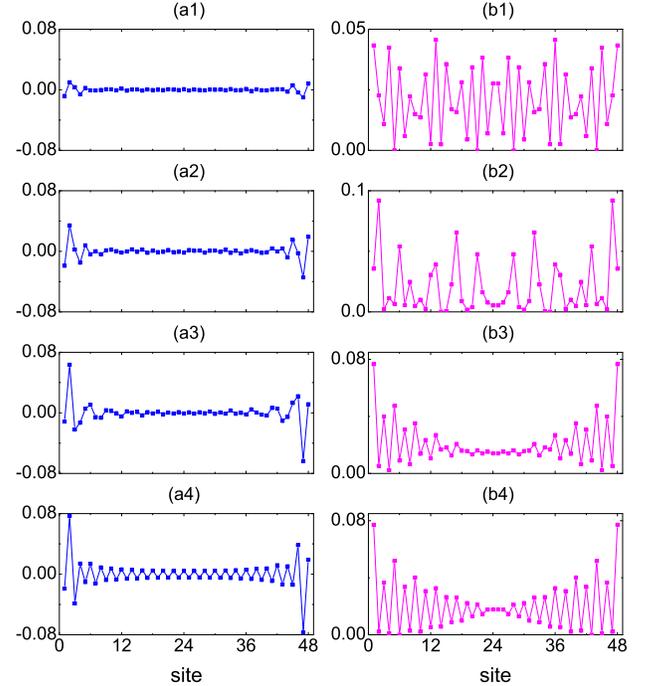}
\caption{(Color online) (a1) to (a4) denote the spatial distributions of the transverse spin polarizations for the normal superfluid phase, LE phase, TSF phase and metal phase under OBC.. (b1) to (b4) denote the normalized single-particle excitation $C(i)$ by adding a spin-down particle in the $N$-particle ground state. Here, $\alpha=0.1$, $0.35$, $0.46$ and $0.6$ for (a1)-(a4) (and (b1)-(b4)), respectively, from top to bottom.}
\end{figure}

To remove out any doubt on the existence of TSF phase in the present system, we consider the boundary effect under OBC. The appearance of edge state under OBC is usually considered to be a hallmark of non-trivial topological properties. For the TSF state in the present number conservation system, though the topological degeneracy no longer exists, there still exist some non-trivial properties for the TSF phase. First, we consider the spatial distributions for the transverse spin polarizations: $\langle \sigma_x(i) \rangle = \langle c_{i\uparrow}^{\dag}c_{i\downarrow}+c_{i\downarrow}^{\dag}c_{i\uparrow} \rangle$, as shown in Fig.5(a1)-(a4). For the normal superfluid state (a1) and LE state (a2), the transverse polarization is distributed in the whole chain while for the TSF state (a3), we observe the accumulation of transverse polarizations at two ends of the chain. For metal phase [Fig.5(a4)], the transverse polarization shows usual staggered distribution. Next, we consider the single-particle excitations by adding a spin-down particle on the chain. For this purpose, we directly calculate the matrix elements between the ground state with particle number $N$ and $N+1$. This quantity directly reveals the distribution of single-particle excitation for the many-body ground state. Define $C(i)=|\langle N+1|C_{i\downarrow}^{\dag}|N\rangle|^2$, where $|N\rangle$ denotes the ground state with $N$ particles. As illustrated in Fig.5(b1)-(b4), for the normal superfluid state (b1) and LE state (b2), the added particle distributes in the whole chain, while for the TSF state, it mainly distributes around the boundary. 
For the metal phase, we still observe clear Friedel oscillations as expected for the metal phase.

{\it Summary.-} In summary, based on DMRG calculations, we have studied exotic pairing states of the SOC Fermi gas in 1D optical lattice with attractive interaction. The interplay between SOC, Zeeman field and attractive interaction induces a series of phases transitions between different pairing states. These different pairing states can be characterized by the pairing correlations in momentum space. We also find the existence of TSF state in the phase diagram of the quarter-filling 1D attractive Fermi lattice system with SOC in a numerically exact way. The TSF phase exists at moderate magnetic field and SOC strength. This non-trivial topological phase can be characterized by its gapless single-particle excitation and nontrivial edge distributions of the transverse spin polarization and single-particle excitation.

{\it Acknowledgment.-} We thank Fuzhou Chen for helpful discussions on the DMRG algorithm. This work has been supported by NSF of China under Grants No. 11425419, No. 11174115, and No. 11325417.

\end{document}